\documentclass[final,5p,times,twocolumn]{elsarticle}

\usepackage{amssymb}
\usepackage[hyphens]{url}
\usepackage{hyperref}
\usepackage{xcolor}
\usepackage{ulem}

\journal{Applied Surface Science}

\begin{document}

\begin{frontmatter}



\title{Observation of a two-dimensional electron gas at CaTiO$_3$ film surfaces}


\author[label1,label2]{Stefan~Muff}
\author[label1,label2]{Mauro~Fanciulli}
\author[label1,label2]{Andrew~P.~Weber}
\author[label2]{Nicolas Pilet}
\author[label2,label3]{Zoran~Risti\'{c}}
\author[label2,label4]{Zhiming~Wang}
\author[label2]{Nicholas~C.~Plumb}
\author[label2,label5]{Milan~Radovi\'{c}}
\author[label1,label2]{J.~Hugo~Dil}

\address[label1]{Institut de Physique, \'{E}cole Polytechnique F\'{e}d\'{e}rale de Lausanne, 1015 Lausanne, Switzerland}
\address[label2]{Swiss Light Source, Paul Scherrer Institut, 5232 Villigen, Switzerland}
\address[label3]{Vin\v{c}a Institute of Nuclear Science, University of Belgrade, 11001 Belgrade, Serbia} 
\address[label4]{Department of Quantum Matter Physics, University of Geneva, 24 Quai Ernest-Ansermet, 1211 Geneva 4, Switzerland}
\address[label5]{SwissFel, Paul Scherrer Institut, 5232 Villigen, Switzerland}

\begin{abstract}

The two-dimensional electron gas at the surface of titanates gathered attention due to its potential to replace conventional silicon based semiconductors in the future. In this study, we investigated films of the parent perovskite CaTiO$_3$, grown by pulsed laser deposition, by means of angular-resolved photoelectron spectroscopy. The films show a c(4$\times$2) surface reconstruction after the growth that is reduced to a p(2$\times$2) reconstruction under UV-light. At the CaTiO$_3$ film surface, a two-dimensional electron gas (2DEG) is found with an occupied band width of 400~meV. With our findings CaTiO$_3$ is added to the group of oxides with a 2DEG at their surface. Our study widens the phase space to investigate strontium and barium doped CaTiO$_3$ and the interplay of ferroelectric properties with the 2DEG at oxide surfaces. This could open up new paths to tailor two-dimensional transport properties of these systems towards possible applications.

\end{abstract}


\end{frontmatter}



\section{Introduction}
\label{sec:intro}

The discovery of a two-dimensional electronic state at the interface of LaAlO$_3$ and SrTiO$_3$ \cite{Ohtomo:2004} triggered research on other oxide interfaces where similar states were found \cite{Hotta:2007, Perna:2010, Gennaro:2013, Chen:2013}. These two-dimensional states at interfaces of complex oxides give rise to different phenomena such as superconductivity \cite{Reyren:2007, Ueno:2008}, metal-insulator transitions \cite{Theil:2006, Cen:2008} or magnetism \cite{Brinkman:2007}. 
More recently, a two-dimensional electron gas (2DEG) was also found on clean SrTiO$_3$ and KTaO$_3$ (001) surfaces \cite{Santander:2011, Meevasana:2011, King:2012, Santander:2012, Plumb:2014}. These states at the vacuum interface can, in contrast to the burried interface states, be more easily probed by angular-resolved photoelectron spectroscopy (ARPES) in the UV-range, revealing their band structure in reciprocal space. It was shown by spin-resolved ARPES, that the 2DEG at the surface of SrTiO$_3$ exhibits a Rashba-like spin splitting of approximately 100 meV, likely enhanced due to the presence of (anti)ferroelectricity and magnetic order at the sample surface \cite{Santander:2014}. 
The strong electron-phonon coupling of the TiO$_2$ surface \cite{Moser:2013, Wang:2016}, which depends on carrier density, is most likely responsible for a drastic rise of the superconducting transition temperature of a monolayer FeSe deposited on top \cite{Lee:2014, Rebec:2017}.
The variety of observed properties makes these oxide-based two-dimensional states an ideal platform to explore new functionalities and possible ways towards device application in the future.

CaTiO$_3$ is the very first discovered perovskite of the transition metal oxide (TMO) family and is thus closely related to the members recent studies focus on. Like SrTiO$_3$, KTaO$_3$ and TiO$_2$ (all compounds shown to host a 2DEG at their surface) CaTiO$_3$ is classified as an incipient ferroelectric or quantum paraelectric material, meaning that it is very close to a ferroelectric phase \cite{Lemanov:1999}. Intermixtures of SrTiO$_3$, BaTiO$_3$ and CaTiO$_3$ form a rich phase diagram, especially regarding the ferroelectric properties, exhibiting para-, ferro- and antiferro-electric phases \cite{Bednorz:1984, Ranjan:2001, Yamamoto:2007}. Pure, crystalline CaTiO$_3$ undergoes two phase transitions at elevated temperatures; from orthorhombic to tetragonal at 1512~K and from tetragonal to cubic at 1635~K \cite{Yashima:2009}. According to band structure calculations for the orthorhombic and cubic crystal lattice the band gap is 2.43~eV or 2.0~eV, respectively \cite{Fan:2015, Tariq:2015}. In today's electronics, CaTiO$_3$ is widely used as a ceramic and as rare-earth doped phosphor with excellent luminescence properties.

In this work, films of 20 unit cells CaTiO$_3$ grown by pulsed laser deposition (PLD) on Nb:SrTiO$_3$ substrates were studied by UV-ARPES and X-ray photoelectron spectroscopy (XPS). Our low-energy electron diffraction (LEED) measurements show that the surface of the CaTiO$_3$ films reconstruct while XPS indicates a TiO$_2$ terminated surface. In addition, observed surface plasmon loss features in the region of the Ti 2\textit{p} core levels suggest the presence of metallic states at the surface of the films. Using ARPES, we found that these metallic states show a purely two-dimensional dispersion with a band width of $\approx$400~meV. Folded bands are visible as an effect of the surface reconstruction. In contrast to SrTiO$_3$ where the mixture of two- and three-dimensional states is observed \cite{Plumb:2014}, this 2DEG is the only metallic state present at the surface. Therefore the CaTiO$_3$ surface states yield easy access to directly manipulate the two-dimensional transport properties of this system by surface structure or gating.   
Furthermore, with the ferroelectricity introduced in Sr$_x$Ca$_{1-x}$TiO$_3$ this is a promising material to investigate the influence of ferroelectricity and the connected electric fields on the 2DEG at the surface of perovskites. 

\begin{figure}[b!]
	\centering
		\includegraphics[width=0.5\textwidth]{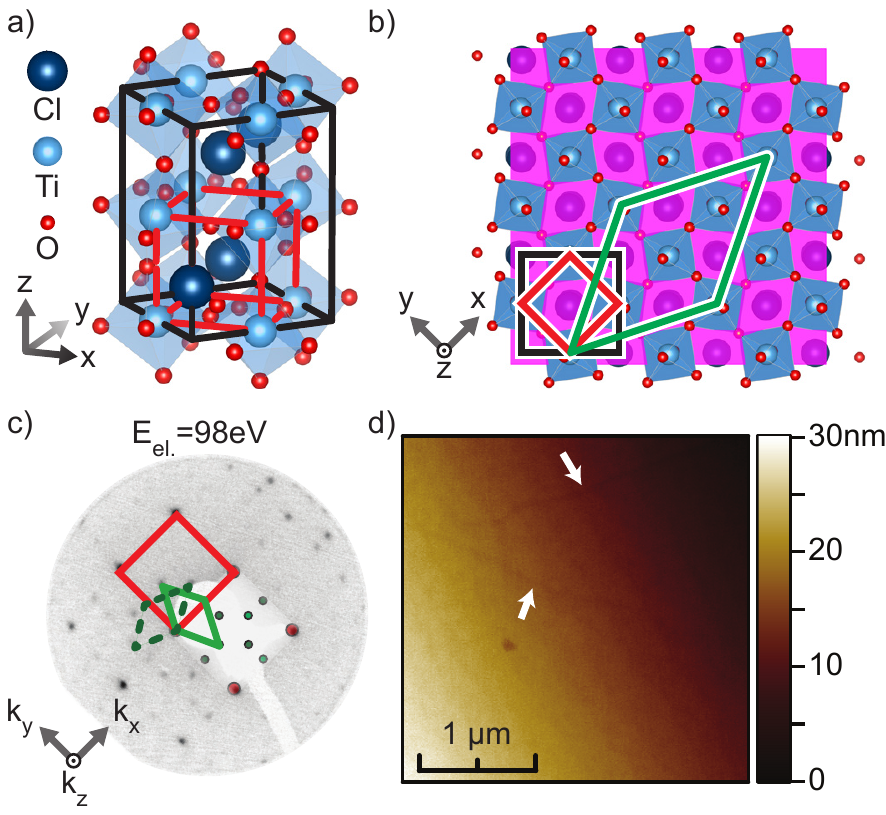}
	\caption{a) Orthorhombic unit cell of bulk CaTiO$_3$ (black) with the simplified pseudo-cubic unit cell (red). b) Crystal surface of TiO$_2$ terminated bulk CaTiO$_3$ in (001) direction with the orthorhombic (black) and pseudo-cubic (red) surface unit cell and the observed c(4$\times$2) reconstruction with respect to the pseudo-cubic lattice (green). c) LEED image with the marked pseudo-cubic Brillouin zone (red) and c(4$\times$2) reconstruction (green) and the 90$^{\circ}$ rotated domain (green dashed). White square inset shows an overlay with the calculated LEED spots for a c(4$\times$2) reconstructed surface. d) AFM topography of the surface. Arrows indicate the domain walls.}
	\label{fig:crystal_leed}
\end{figure}

\section{Materials and experimental method}
\label{sec:exp}

The CaTiO$_3$ films of 20 unit cell thickness used for this study where grown by PLD on commercial TiO$_2$ terminated SrTiO$_3$ (001) substrates with a niobium doping of 0.5 wt$\%$ (Twente Solid State Technology BV). The growth was performed at a substrate temperature of 680$^{\circ}$~C in partial oxygen pressure of 5$\times10^{-5}$ mbar. The growth process and film thickness was monitored by reflection high-energy electron diffraction.
The prepared films were \textit{in-situ} transferred to the experimental station at the Surface and Interface Spectroscopy beam line of the Swiss Light Source at the Paul Scherrer Institut under ultra high vacuum (UHV) conditions and measured without further treatment. 
The sample was held at a temperature of 20~K in pressures better than 8$\times10^{-11}$~mbar during the measurements. Photoemission spectra (XPS and ARPES) were taken using a Scienta R4000 hemispherical electron analyzer and circular polarized synchrotron light. LEED patterns were obtained at 20~K before the ARPES measurements. The atomic force microscopy (AFM) topography was measured at the NanoXAS beam line of the Swiss Light Source at the Paul Scherrer Institut with the sample at room temperature in UHV environment.

The orthorhombic unit cell of bulk crystalline CaTiO$_3$ has lattice parameters of $a=5.367$~\AA, $b=7.644$~\AA and $c=5.444$~\AA~\cite{Kay:1957}. An approximate representation of the orthorhombic unit cell can be made by a pseudo-cubic unit cell as marked in Fig.\ref{fig:crystal_leed}a). The lattice parameters of the pseudo-cubic unit cell $a/\sqrt{2} \approx b/2 \approx c/\sqrt{2}\approx 3.822$~\AA~are similar to cubic SrTiO$_3$ with a lattice mismatch of approximately 2\%. 

In LEED we can identify the primary diffraction spots corresponding to the pseudo-cubic unit cell. Further we observe spots indicating a c(4$\times$2) surface reconstruction of the pseudo-cubic lattice with domains rotated 90$^{\circ}$ with respect to each other (see Fig.\ref{fig:crystal_leed}b) and c)). The (1$\times$1) TiO$_2$ terminated surface at the vacuum interface of TMO perovskites might be unstable due to the unshared oxygen atom of the TiO$_2$ polyhedron sticking out of the surface. 
Of the surface reconstructions reported for the closely related SrTiO$_3$ system, c(4$\times$2) reconstruction has also been observed \cite{Castell:2002, Erdman:2003, Iwaya:2010, Zhu:2012}.

The AFM topography in Fig.\ref{fig:crystal_leed}d) shows that the films are of low roughness and follow the substrate steps with a terrace size of approximately 200~nm. However, the AFM measurements do not have the resolution required to observe the surface reconstruction. The observed presence of domain walls is a further indication of the existence of multiple rotated domains corroborating the LEED data.

\section{Results and discussion}
\label{sec:res}

\begin{figure*}[t!]
	\centering
		\includegraphics[width=0.70\textwidth]{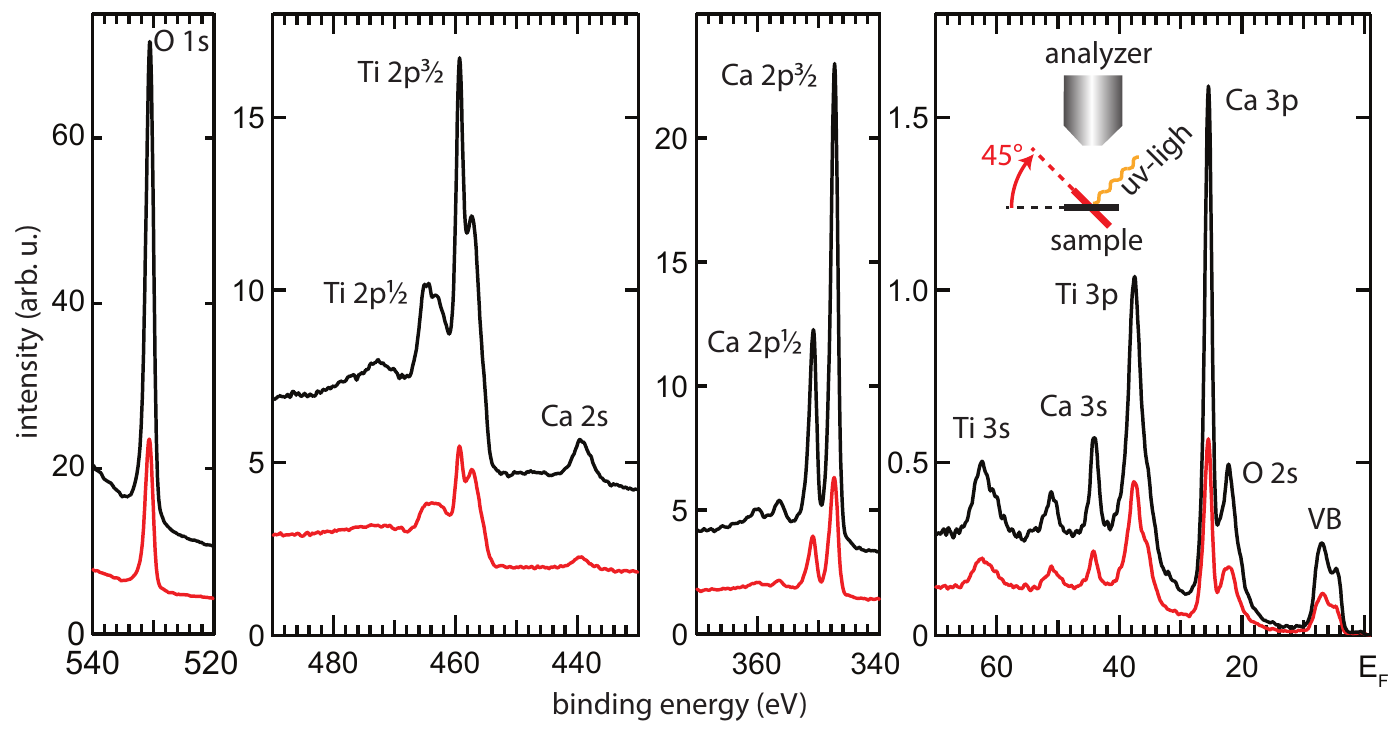}
	\caption{Core level intensity obtained with $h\nu$ = 600 eV at normal emission (black) and 45$^{\circ}$ rotated (red) as illustrated in the setup sketch.}
	\label{fig:xps}
\end{figure*}

The XPS spectrum of the films in Fig.\ref{fig:xps}, measured with a photon energy of $h\nu$~=~600~eV, shows clear signatures of the expected calcium, titanium and oxygen core levels with no detectable contamination. Comparing the spectra taken with the sample surface normal to the analyzer to the more surface sensitive measurement taken at an angle of 45$^{\circ}$ between the sample normal and the analyzer axis (see sketch inset in Fig.\ref{fig:xps}) we can confirm the TiO$_2$ termination of the grown films. This termination of the film surface is expected due to the TiO$_2$ termination of the SrTiO$_3$ substrate \cite{radovic:2009}. When comparing the peak areas ($A_i$) after background subtraction the ratio $A_{Ca~2\textit{p}}/A_{Ti~2\textit{p}}$ of 0.75 at normal emission is significantly higher than the ratio of 0.65 measured at an emission angle of 45$^{\circ}$. 

All the titanium peaks show a shoulder towards lower binding energy, indicating the existence of titanium atoms with different valency. The increase of the surface located Ti 3$^+$ shoulder is a light induced effect commonly observed in this class of materials \cite{Plumb:2014}. The appearance of Ti 3$^+$ ions is likely linked to a distortion of the TiO$_2$ octahedron, for example due to the creation of oxygen vacancies in the surface region and/or a structural rearrangement and buckling of the surface layers.

\begin{figure}[b!]
	\centering
		\includegraphics[width=0.48\textwidth]{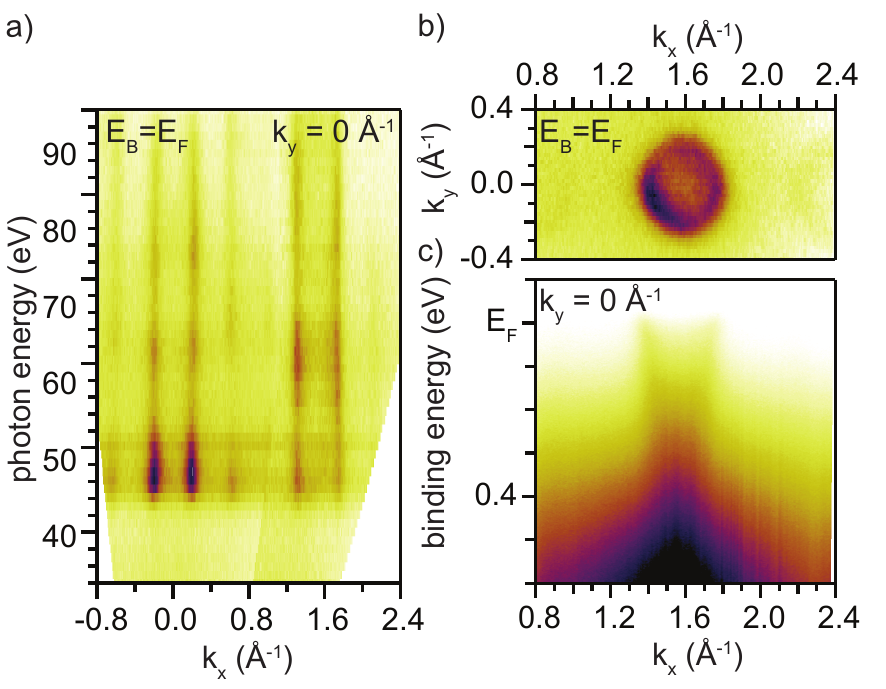}
	\caption{a) Dispersion at the Fermi energy as a function of photon energy and $k_x$. b) Fermi surface measured at $h\nu$ = 60 eV and c) band structure along the high symmetry direction $\overline{\Gamma \mbox{X}}$.}
	\label{fig:hv_60eV}
\end{figure}

The Ti 2\textit{p} as well as the Ca 2\textit{p} core levels show plasmon loss peaks in their shake-up tail with an energy loss of 13.2~eV for titanium and 9~eV for calcium. Plasmon loss peaks with this loss energy of the Ti 2\textit{p} core levels have been observed in other perovskites. The measured plasmon energy corresponds to surface plasmons present in TiO$_2$ where the plasmons are trapped at the interface of the metallic surface and the dielectric bulk due to the sudden change in dielectric constant. \cite{Sen:1976, deBoer:1984, Bocquet:1996, Bahadur:2010}  

Consequently, we also expect metallic states to be present at the surface of our CaTiO$_3$ films. Indeed, the ARPES measurements in Fig.\ref{fig:hv_60eV} show an electron-like surface state. The scan  over a wide range of photon energies in Fig.\ref{fig:hv_60eV}a) shows no dispersion of these states with out-of-plane momentum, verifying their two-dimensional nature. In contrast to the well-studied metallic states present at the surface of SrTiO$_3$ (001) and KTaO$_3$ (001) \cite{Santander:2011, Meevasana:2011, King:2012, Santander:2012, Plumb:2014} we have no indication of three-dimensional features, making the 2DEG  the only states contributing to the metallicity. Similar to the other perovskites the spectral intensity of the 2DEG at the CaTiO$_3$ surface increases under UV-irradiation. This is attributed to light induced surface rearrangements and induced carriers \cite{Plumb:2014}. 


The circular Fermi surface of $\Gamma_{\left( 100 \right)}$ is depicted in Fig.\ref{fig:hv_60eV}b) and the corresponding clear free-electron-like parabolic band along the high symmetry direction $\overline{\Gamma \mbox{X}}$ in Fig.\ref{fig:hv_60eV}c). The in-plane momentum $g\approx 1.57$~\AA$^{-1}$ at the ring center, corresponding to the momentum of $\Gamma_{\left( 100 \right)}$, is equal to a lattice parameter of $a\approx 4$ \AA. 
This is in good agreement with the lattice parameter of the pseudo-cubic unit cell of CaTiO$_3$ and the SrTiO$_3$ substrate.
Also clearly visible  in Fig.\ref{fig:hv_60eV}a) is the intensity at the Fermi energy of an additional, folded parabola between $\Gamma_{\left( 000 \right)}$ and $\Gamma_{\left( 100 \right)}$ due to the surface reconstruction observed also in LEED as described in section \ref{sec:exp}. Similar band folding has been observed for the (1$\times$4) reconstructed anatase TiO$_2$ films \cite{Wang:2017}.

\begin{figure*}[t!]
	\centering
		\includegraphics[width=0.70\textwidth]{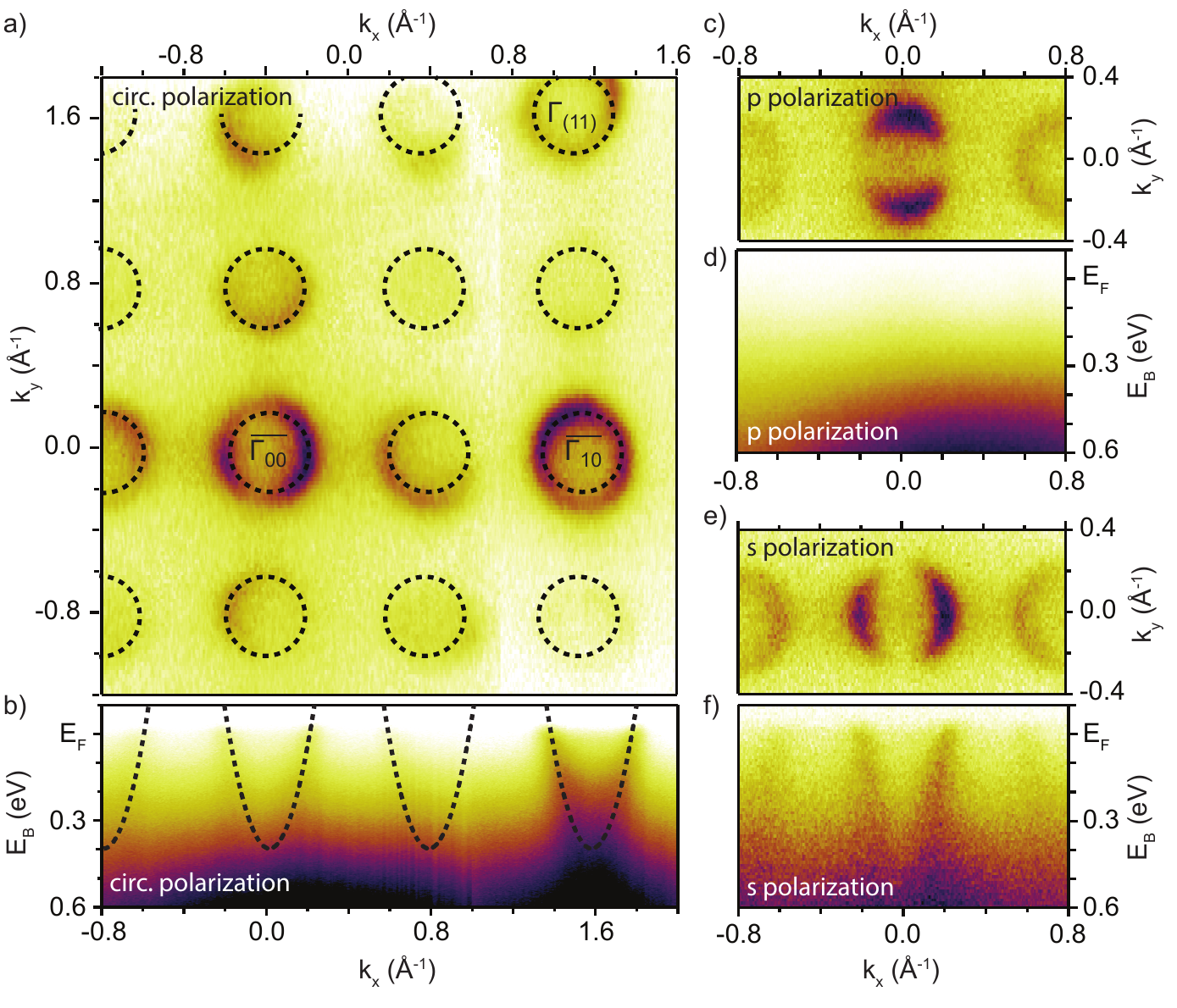}
	\caption{a) Fermi surface at $h\nu$ = 80eV for circular polarized light. Dashed black circles mark the expected Fermi surface due to the p(2$\times$2) reconstruction. b) Band dispersion along the $\overline{\Gamma \mbox{X}}$ direction with indicated positions of the free electron like parabolas. c) Fermi surface and d) band structure for p-polarized light of $\overline{\Gamma_{00}}$. e) Fermi surface and f) band structure for s-polarized light.}
	\label{fig:cube80eV}
\end{figure*} 

The band structure of the 2DEG in Fig.\ref{fig:hv_60eV}c) and \ref{fig:cube80eV}b) can be fitted with a free-electron-like parabola yielding an effective mass of $m^* \approx 0.39 m_e$, a Fermi momentum of $k_F \approx 0.20$ \AA$^{-1}$, a Fermi velocity of $v_F \approx 6.3\times10^5$~m/s, and a band minimum at a binding energy of $E_b\approx$400~meV. This corresponds to a charge carrier density per parabola of $6.4 \times 10^{13}~cm~^{-2}$ or $0.1~e^-/a^2$ with $a=3.822$~\AA. This charge carrier density is similar to SrTiO$_3$ \cite{Meevasana:2011, Plumb:2014} while the band width is significantly higher and the effective mass much lower than for SrTiO$_3$ and KTaO$_3$. 
The ARPES measurements with s- and p-polarized light in Fig.\ref{fig:cube80eV}(c-f) confirm the $xy$-symmetry of the 2DEG with no indications of bands with $xz$- or $yz$-symmetry. The 2DEG thus consists of the Ti 3\textit{d}$_{xy}$ bands splitted from \textit{d}$_{xz}$/\textit{d}$_{yz}$ by crystal field splitting and partially filled due to surface band bending and light induced carriers. 

With the absence of the Ti 3\textit{d}$_{xz}$ and 3\textit{d}$_{yz}$ bands and the two-dimensional Ti 3\textit{d}$_{xy}$ bands at relatively high binding energies, the splitting between the \textit{d}$_{xy}$ and \textit{d}$_{xz}$/\textit{d}$_{yz}$-bands has to be large, at least of the size of the observed bandwidth of $\approx$400~meV. This splitting is considerably larger than the 240~meV measured for SrTiO$_3$ \cite{Plumb:2014} but smaller than for TiO$_2$ anatase where 1~eV is reported \cite{Moser:2013, Wang:2016}. For the orthorhombic oxide LaAlO$_3$ a comparable noncubic crystal field splitting of 120~meV to 300~meV for the \textit{$t_{2g}$} sub shell is reported \cite{Haverkort:2005}.
However, there is no detectable additional splitting of the Ti 3\textit{d}$_{xy}$ band as observed for SrTiO$_3$ \cite{Santander:2014}. Comparing SrTiO$_3$ to CaTiO$_3$ the increased rotation of the TiO$_3$ octahedron in the later due to the orthorhombicity will likely reduce the local electric fields as observed in other perovskites \cite{Sando:2014}. The resulting weak polarization field at the surface could be the reason that the splitting is too small to be observed in our data.

The results of the fitting are indicated in Fig.\ref{fig:cube80eV}a) and b) showing the circular Fermi surface composed by parabolic bands for the primary $\Gamma$-points as well as for the reconstructed $\Gamma$-points. Along the $\overline{\Gamma \mbox{X}}$ direction, the Fermi surfaces and parabolic bands corresponding to the folded $\Gamma$-points, which are present as a result of the reconstruction, are clearly visible in the data. However their intensity is weaker than the signal of the 2DEG at the primary $\Gamma$-points. In contrast to the folding along the high-symmetry direction, the $\Gamma$-points offset by $1/4\cdot g$ in $k_y$ direction are not present in the data. A possible reason for this is a change of the reconstruction from c(4$\times$2) either to a combination of (2$\times$1) and c(2$\times$2) or more likely to p(2$\times$2) under irradiation with UV-light. Since we observe an increasing intensity of the 2DEG as well as the described formation of a low binding energy shoulder on the titanium core levels under UV-light, a change of reconstruction under light due to the deposited energy is plausible.

\section{Summary}
\label{sec:conc}

In conclusion, we have revealed the existence of metallic states at the surface of CaTiO$_3$ films consisting solely of a 2DEG. The 2DEG has a band width of $\approx$400~meV, indicating a large splitting between the unoccupied Ti 3\textit{d}$_{xz}$/\textit{d}$_{yz}$ bands and the two-dimensional 3\textit{d}$_{xy}$ bands. Due to its metallicity, the surface also hosts plasmons visible as loss peaks in the XPS data. The bands are folded according to the surface reconstruction that is likely changed from c(4$\times$2) to p($2\times$2) under UV irradiation. 
Due to the lack of higher-dimensional conducting channels and the affinity of the system to adapt to its surface structure, various paths open up to directly manipulate the surface states. This manipulation may give direct access to the transport properties of the system and its coupling to overlayers. 
With the possibility to induce ferroelectricity into the quantum paraelectric materials CaTiO$_3$ and SrTiO$_3$ by mutual doping, the phase space is open to probe the effect of ferroelectricity on the 2DEG hosted by both of these materials.

\section*{Acknowledgments}
\label{sec:ack}

We acknowledge financial support from the Swiss National Science foundation Project No. PP00P2\_144742/1. 
\newline





\bibliographystyle{elsarticle-num} 



\end{document}